\documentclass[12pt,a4paper]{article}
\usepackage{bezier}
\usepackage{epsfig}
\usepackage{amssymb}
\usepackage{graphicx}
\usepackage{psfrag}

\newcommand{\nc}{\newcommand}
\nc{\be}{\begin{equation}}
\nc{\ee}{\end{equation}}
\nc{\bea}{\begin{eqnarray}}
\nc{\eea}{\end{eqnarray}}
\newcommand{\av}[1]{\langle #1\rangle}

\nc{\eqn}[1]{{(\ref{#1})}}
\nc{\cA}{{\cal A}}
\nc{\cB}{{\cal B}}
\nc{\cC}{{\cal C}}
\nc{\cD}{{\cal D}}
\nc{\cE}{{\cal E}}
\nc{\cF}{{\cal F}}
\nc{\cG}{{\cal G}}
\nc{\cH}{{\cal H}}
\nc{\cI}{{\cal I}}
\nc{\cJ}{{\cal J}}
\nc{\cK}{{\cal K}}
\nc{\cL}{{\cal L}}
\nc{\cM}{{\cal M}}
\nc{\cN}{{\cal N}}
\nc{\cO}{{\cal O}}
\nc{\cP}{{\cal P}}
\nc{\cQ}{{\cal Q}}
\nc{\cR}{{\cal R}}
\nc{\cS}{{\cal S}}
\nc{\cT}{{\cal T}}
\nc{\cU}{{\cal U}}
\nc{\cV}{{\cal V}}
\nc{\cW}{{\cal W}}
\nc{\cX}{{\cal X}}
\nc{\cY}{{\cal Y}}
\nc{\cZ}{{\cal Z}}
\nc{\tr}{{{\rm tr}\,}}

\nc{\bk}{{{\bf k}}}
\nc{\bx}{{{\bf x}}}

\nc{\simo}[1]{{\stackrel{#1}{\simeq}}}
\nc{\geqo}[1]{{\stackrel{#1}{\geq}}}
\nc{\geo}[1]{{\stackrel{#1}{>}}}
\nc{\guo}[1]{{\stackrel{#1}{\succ}}}

\nc{\rbo}{\raisebox}
\nc{\RR} {\rangle \! \rangle}
\nc{\LL} {\langle \! \langle}
\nc{\rmi}[1]{{\mbox{\small #1}}}
\nc{\eq}{eq.~}
\nc{\nr}[1]{(\ref{#1})}
\nc{\ul}{\bf }
\nc{\mc}{\multicolumn}
\nc{\todo}[1]{\par\noindent{\bf $\rightarrow$ #1}}

\nc{\cu}{{\cal u}}

\begin{document}

\begin{flushright}
SPhT-00/012 \\
BI-TP 00/07\\
HD-THEP-00-05
\end{flushright}

\begin{center}	
\large{\bf High Temperature 3D QCD: Dimensional Reduction at Work}
\end{center}

\begin{center}	
P. Bialas$^{1,2,a}$, A. Morel$^{3,b}$, B.Petersson$^{1,c}$, K. Petrov$^{1,d}$ 
and T. Reisz$^{3,4,e}$
\end{center}

\centerline{$^{1}$Fakult\"at f\"ur Physik, Universit\"at Bielefeld} 
\centerline{P.O.Box 100131, D-33501 Bielefeld, Germany}
\vspace{0.3cm}
\centerline{$^{2}$Inst. of Comp. Science, Jagellonian University} 
\centerline{33-072 Krakow, Poland}
\vspace{0.3cm}
\centerline{$^3$Service de Physique Th\'eorique de Saclay, CE-Saclay,} 
\centerline{F-91191 Gif-sur-Yvette Cedex, France} 
\vspace{0.3cm}
\centerline{$^4$Institut f\"ur Theor.~Physik, Universit\"at Heidelberg}
\centerline{Philosophenweg 16, D-69120 Heidelberg, Germany}
\vspace{0.3cm}

\begin{abstract} \normalsize
We investigate the three-dimensional SU(3) gauge theory at finite
temperature in the framework of dimensional reduction.  The large
scale properties of this theory are expected to be conceptually more
complicated than in four dimensions.  The dimensionally reduced action
is computed in closed analytical form.  The resulting effective
two-dimensional theory is studied numerically both in the electric and
magnetic sector.  We find that dimensional reduction works excellently
down to temperatures of 1.5 times the deconfinement phase transition
temperature and even on rather short length scales.  We obtain strong
evidence that for ${\rm QCD}_3$, even at  high temperature the
colour averaged potential is represented by the exchange of a single
state, at variance with the usual Debye screening picture involving a
pair of electric gluons.
\end{abstract}

\vfill

\noindent
$^a$pbialas@physik.uni-bielefeld.de \\
$^b$morel@spht.saclay.cea.fr \\
$^c$bengt@physik.uni-bielefeld.de \\
$^d$petrov@physik.uni-bielefeld.de \\
$^e$t.reisz@thphys.uni-heidelberg.de \\

\newpage

\section{\bf Introduction } 
\vspace{0.5cm}

Dimensional reduction is a powerful technique
to study field theories at high temperature. In the Euclidean
formulation such theories are defined in a volume with one
compact dimension of extent $1/T$. As the temperature T becomes
large, one may expect that the non-static modes in the
temperature direction can be neglected, and that one is left
with a theory in one less dimension \cite{ginsparg,appelquist}.
This naive reduction is, however, only exact in the classical
limit. In general, one has to integrate over the nonstatic
modes to obtain the effective action \cite{nadkarni,landsman,thomas}.
As was shown in Refs
\cite{nadkarni,thomas}
the effective action for the long distance phenomena will, however
only contain a limited number of local terms at sufficiently high
 temperature, because the integral over the
nonstatic modes does not contain infrared divergencies.
Furthermore the coefficients can be 
determined from a perturbative expansion of the integral
over the non-static modes. This effective action is then expected
to describe correctly the infrared behaviour of the full
theory.

This method was first successfully applied to 
lattice gauge theories in
Refs \cite{thomas,su2,su3}. In particular the 
screening mass as defined from the correlation between
Polyakov loops, which describes the colour averaged
potential at finite temperature, was calculated.
The renormalized coupling constant, used in the reduction, 
was defined from a perturbative calculation of this potential
at small distances.
In the application to
pure lattice gauge theories it was possible to compare the
screening length in the reduced theory with that calculated in the full theory. It was
found that the dimensional reduction gave good results in the
gluon plasma phase down to $T/T_c \approx 2$, where $T_c$ is the
transition temperature between the confined and the plasma phase.

The real challenge is of course the calculation in full QCD.
Here the fermions can be integrated out explicitely, because
they have no static modes, and they only change the coefficients
in the effective action. A calculation of the screening lengths
was performed in \cite{qcd}. A summary can be found in
\cite{lacock}.

Later the formalism was adapted to and successfully used to
investigate the electroweak phase transition in \cite{kajantie0}. It
was also employed to determine the chromoelectric screeening mass in
QCD over a large range of values of the temperature
\cite{kajantie1,kajantie2,kajantie3}. In these articles another
operator was used to define a screening mass. Also another
renormalization scheme was used, the $\overline{\mbox{MS}}$ scheme,
leading to a renormalized squared coupling around three times larger in
the similar temperature range. One may thus expect higher order
corrections to be more important. Further evidence for dimensional
reduction in the region of a few times $T_c$ have been found in
\cite{datta}, \cite{karschpetr} and \cite{hart}.

It is, of course, very interesting if several quantities can
be calculated in the same scheme. In particular quantities depending
on the chromomagnetic sector, which have infrared singularities
in perturbation theory that
cannot be determined so far by
resummation, need a non-perturbative
approach. One gauge independent operator, which has been
studied is the spatial string tension. However, it is still an open
question what the corresponding observable is in the reduced model. 
In Ref.~\cite{space} 
it was shown that this string tension
increases with the temperature in the deconfined phase. Later
it was compared with the three-dimensional naively reduced theory, and
a good agreement was found \cite{karsch}, using, however, 
the bare coupling constant,
which even on moderately sized lattice is different from the 
renormalized one employed in Ref. \cite{thomas} - \cite{qcd}.

In the present article we want to show more precisely how well
dimensional reduction works, both for correlations between Polyakov
loops and for the spatial string tension, in the case of three
dimensional SU(3) gauge theory at high temperature. The effective
model in this case is a two-dimensional Higgs model, with the
Higgs field in the adjoint representation. The interest of this
calculation is that  the three-dimensional theory has many
properties in common with the four-dimensional one, but the numerical
simulations become much simpler, and one can obtain much more precise
information on the quality of dimensional reduction. However,
the infrared behaviour of perturbation theory is worse, so that even the
leading perturbative definition of the Debye mass is not well
defined.
Three dimensional gauge theories at finite temperature
have been considered earlier by D'Hoker \cite{dhoker}.

In this article the three-dimensional model is investigated through numerical
lattice simulations at finite temperature. Dimensional
reduction is performed by perturbation theory in the static time
averaged Landau gauge, using the lattice regularization, when 
necessary. The effective two dimensional model is also investigated
by lattice simulations, which automatically include the non
perturbative features of the model. A detailed comparison
of correlations between Polyakov loops and of the spatial string
tension is presented.
Because the spatial dimension is two, we expect that the
infrared properties of the theory are even more complicated
than in $3+1$ dimensions. 
In particular, the Debye screening picture does not 
necessarily translate to QCD${}_3$.

Our paper is organized as follows. In the next section we introduce 
the lattice implementation of 3-dimensional SU(3) QCD at high 
temperature and derive the associated dimensionally reduced
model. Special care is taken in defining the 
self couplings of the adjoint scalar of the 2D action in the limits 
appropriate to the large temperature behaviour of QCD${}_3$ in the 
continuum (scaling limit). Section 3 is devoted to the 
corresponding numerical simulations in 2 and 3 dimensions, and  to
the measurements of Polyakov loop correlations 
and spatial Wilson loops. The results are presented and
discussed in Section 4. We show that the Polyakov loop correlations measured in
(2+1)D or through their 2D reduction agree at large distances, and
remain close to each other down to quite small distances, the
more so the temperature is high. This finding gives
quantitative support to the statement that, for static
observables, the reduced action correctly describes the (2+1)D
properties in the limit where the momenta are small compared to
temperature. The decay of these correlations with spatial distance
is not compatible with the Debye screening expected from the exchange 
of two electric gluons, but strongly favours the existence of an isolated
excitation whose mass is measured.
We also show that, although the Wilson loops are
non-static operators, the (spatial) string tension in (2+1)D is close to
that measured in 2D, which itself differs only weakly from the known
pure gauge 2D string tension. We summarize and conclude in a last
section.

\section {\bf Lattice Action in 3D and its Reduction to~2D  }
\subsection{\bf Lattice Definitions}

In this subsection we give the definitions pertinent to the
simulation of SU(3) gauge theory in $2+1$ dimensions, at
finite temperature.

The gauge field action is taken to be the Wilson
action on a cubic lattice $\Lambda_3$, namely
\be	
   S_W^3(U) = \beta_3 \sum_{x\in\Lambda_3} \sum_{\mu<\nu=0}^2
    \biggl( 1 - \frac{1}{3} \Re \,\tr U(x;\mu) U(x+a\widehat\mu;\nu)
     U(x+a\widehat\nu;\mu)^{-1} U(x;\nu)^{-1} 
    \biggr). \label{wilson}
\ee
The axis $\mu =0$ is along the Euclidean time or temperature
direction, and $a$ is the lattice spacing.
An SU(3) gauge group element $U(x;\mu)$ is related to the gauge fields
$A_{\mu}^d(x)$, $d\in [1,8]$, by
\bea
   && U(x;\mu) = \exp[i\,A_\mu(x)],  \\
   && A_\mu(x) = \sum_{d=1}^{8} A_\mu^d(x)\lambda^d
    , \quad \tr \lambda^d \lambda^e =  \frac {1}{2} \delta_{de}.
\eea
The $\lambda$'s denote the $SU(3)$ algebra generators. The gauge fields
are dimensionless.
The lattice coupling $\beta_3$ is related to
the conventional bare gauge coupling $g_3$ of perturbative QCD by
\be
 \beta _3=\frac {6}{ag_3^2}.  \label {beta}\\
\ee
The theory is super-renormalizable, $g_3^2$, which has 
dimension one in energy, is proportional to the renormalized coupling constant
squared, and can thus be used to set the physical scale. 
The 3D lattice action  describing the system at a finite temperature $T$
is obtained by taking for $\Lambda _3$ a cubic lattice $L_s^2\times L_0$, and
\be
T=\frac {1}{aL_0}. \label {temp}
\ee
\vspace {0.3cm}

The observables of interest in this work are the {\it spatial} 
string tension, as extracted from the expectation
values of spatial Wilson loops 
formed by links $\vec{\ell}$ joining the sites $x=\vec {x}+x_0\,\widehat{0}$ 
of a rectangle $R=[R_1,R_2]$ at $x_0$ fixed,
\be
W(R_1, R_2)=\frac {1}{3} \tr \prod_{\vec{x} \in R} U(x;\widehat{\ell}), 
\label{wloop}
\ee
and the expectation values of Polyakov loops $L(\vec{x})$
and their connected correlations $P(\vec{x})$
\bea
L(\vec{x})&=&\frac {1}{3}\tr \prod_{x_0=1}^{L_0}
U(\vec{x}+x_0\, a \,\widehat{0},0).  \label{ploop}  \\ 
P(\vec{x})&=&\av{L(\vec{c})\, L^\dagger(\vec{c}+\vec{x})}-|\av{L(\vec{c})}|^2.
\label{plc} 
\eea

\vspace {0.3cm}
We want to study the behaviour at large $T$ of the continuum 
limit of the theory defined by $S_W^3$. Before proceeding,
let us state more precisely what these continuum and large $T$ limits 
mean in terms of the lattice parameters $L_0,\,
\beta _3$ and $L_s$. Given $g_3$ and $T$ we have from (\ref
{beta},\,\ref {temp})
\bea
 \beta _3=\frac {6} {a\,g_3^2},\quad\mbox{and}\quad L_0=\frac {1}{a\,T}.   
\eea
It follows that as $a\to 0$, scaling (constant physics)
corresponds to 
\bea
L_0\to \infty , \quad \beta _3 \to\infty,\qquad 
\tau \equiv \frac {\beta _3}{6L_0}= \frac
{T}{g_3^2}=constant. \label {scaling}
\eea
The dimensionless quantity $\tau$ thus measures temperature in units of
the scale $g_3^2$, high temperature means large $\tau$ values.
From numerical simulations we obtain that $T_c/g_3^2\approx 0.61$ \cite{lego}. 
Also, on the lattice, $L_s$ must be kept much larger than the
largest spatial correlation length in lattice units. Assuming
that this length is of order 
$1/(ag_3^\alpha {T}^{(1-\alpha/2)})$ for some $\alpha>0$, this
together with (\ref {temp}) implies 
\begin{equation} \label{volume_scaling}	
\frac {L_s}{L_0}\gg \left( \frac {\sqrt T}{g_3} \right)^\alpha ={\sqrt 
\tau}^\alpha .
\end{equation}
Note that this ratio increases with temperature.
As we shall see, these considerations, important for the design a 
meaningful numerical investigation of the high temperature phase of 
QCD${}_3$, are also useful to discuss the
adequacy of its 2D reduction to describe at least part of its
large distance physics. 

\subsection {\bf Dimensional Reduction: the Effective 2D Action}

The dimensional reduction means that one should derive an effective
action for the static modes, obtained by keeping the static part of
the action and adding terms coming from a perturbative integration
over the non-static modes. It is therefore convenient to choose
a static gauge, where this separation can be made.

The operators of interest to us are gauge invariant and may be computed in any
gauge. In particular, a static gauge can be realized on the lattice with 
periodic
boundary conditions by choosing $A_0$ independent of the imaginary
time coordinate $x_0$:
\be
A_0(x_0,\vec{x})=A_0(\vec{x}). \label {static} \\
\ee
In such a gauge the Polyakov loops are static operators, unlike
spatial Wilson loops, and read
\be
L(\vec{x})=\frac {1}{3}\tr \exp\bigl [i\,L_0\,A_0(\vec {x})\bigr ].
\label {statl}
\ee

Full gauge fixing, which is necessary for forthcoming perturbative calculations
\cite {petereisz}, is achieved by adding the Landau constraint
\be
\sum_{x_0} \sum_{i=1}^2 \bigl [A_i(x)-A_i(x-a\widehat{i})\bigr
]=0.\label{landau}
\ee
Together with (\ref {static}), this condition constitutes the so-called
static,
time-averaged Landau gauge (STALG, see \cite {curci,STALG}).

At high temperature and in the thermodynamical limit, the well known $Z_3$
symmetry of
the $S_W^3$ action is broken (deconfined phase) and 
the average Polyakov loop $\av{L(\vec{y})}$ is not zero. Its
phase may be either one of the three $3^{rd}$ roots of one, the
corresponding threefold degeneracy can be lifted up by rotating this
phase away, so that the $U$'s are always connected to the unit
matrix.  

In what follows, then, the gauge manifold will be parametrized by
gauge fields $A_\mu^d(x)$ subjected to the constraints (\ref{static},
\ref
{landau}), and perturbation theory, whenever it applies, is
generated by
expanding the 3D action into small fields.

\vspace {0.3cm}

The effective 2-dimensional action is obtained according to standard
techniques (\cite {ginsparg}-\cite {thomas}). It consists in the
classically reduced 3D action, with {\it all fields} restricted to
their static components (no dependence in the $0^{th}$ coordinate),
supplemented by the interactions generated by integrating over the
non-static degrees of freedom which are left over after gauge fixing,
namely the non-static part $A_i^{ns}(x_0,\vec{x})$, i=1,2, of the
gauge fields, constrained by (\ref {landau}). The resulting action
depends on the 2-dimensional SU(3) gauge fields
$A_i^{static}(\vec{x})$ defined by 
\bea 
&& A_i(x) =
A_i^{static}(\vec{x}) \; + \; A_i^{ns}(x_0,\vec{x}), \\ && 
\sum_{x_0}
A_i^{ns}(x_0,\vec{x}) = 0 , 
\eea 
and on $A_0(\vec{x})$, a scalar in 2
dimensions (Higgs field in the adjoint representation). In the
following, the superscript {\it static} on the 2D gauge fields will
be omitted. So defined, this action is of course non polynomial and
non local in $A_0$.

 At large $T$ however, one may truncate its expansion in powers of 
$A(\vec{x})$, adapting the method developed in details in \cite {thomas} for 
the 4D$\to$3D reduction. In the present case, power counting 
and Becchi-Rouet-Stora symmetry of the gauge fixed theory leads to retain
only the 1-loop contribution of the non-static modes to the 
quadratic and quartic $A_0$ self-couplings at external momenta 
$p=0$. 
The contributions of 
monomials of higher degree, or non local in $A_0$ are suppressed by powers of 
$(g_3^2/T)^2$ and $(p/T)^2$, up to $\log$'s. More precise statements will
be made at the end of this section.\\

We write down the explicit result $S_{eff}^2$ of this reduction,
in terms of the parameters $L_0$ and $\beta _3$ of the initial 3D
lattice action ($L_s$ is assumed to satisfy (\ref{volume_scaling})
so that finite size effects are negligible).

 For later convenience, we rescale the scalar field
by setting
\be
 A_0(\vec{x})= \phi(\vec{x}) {\sqrt \frac {6} {L_0\,\beta _3}}, \label
 {phi} \\
\ee
and express $S_{eff}^2$ as a function of $\phi$ and of the 2-dimensional $U$'s.
It contains three pieces, a pure gauge part $S_W^2$, the gauge
covariant kinetic term $S_{U,\phi}$ for $\phi$, and the
self-interaction $S_{\phi}$ induced by integration over the non
static fields:

\be  \label{seff}
S_{eff}^2(U,\phi)=S_W^2+S_{U,\phi}+S_{\phi}.
\ee
Here $S_W^2$ is the Wilson action for $D=2$ obtained by repeating   
$L_0$ times the purely spatial part of $S_W^3$. Hence 
\begin{eqnarray}
   S_W^2 &=& \beta_2 \sum_{\vec{x}\in\Lambda_2} 
    \biggl( 1 - \frac{1}{3} \Re\, \tr U(\vec{x};1) 
      U(\vec{x}+a\widehat 1;2)
     U(\vec{x}+a\widehat 2;1)^{-1} U(\vec{x};2)^{-1} 
    \biggr), \label{sw2}\\
\beta_2&=&\beta_3 L_0. \label {beta2}
\eea
The lattice kinetic term for the (adjoint) field $\phi$ is
obtained by expanding the $S_W^3$ action to second order in
$A_0$, and reads
\bea \label {suphi} 
    S_{U,\phi} &=&\sum_{\vec{x}} \sum_{i=1}^{2}
    \tr \biggl( D_i(U) \phi (\vec{x}) \biggr)^2, \\
     D_i(U) \phi (\vec{x}) &=& U(\vec{x};i) \phi (\vec{x}+a 
     \widehat i) U(\vec{x};i)^{-1} - \phi(\vec{x}). 
\eea
Finally the ``Higgs potential" $S_{\phi}$ is written
\be
   S_{\phi} = \sum_{\vec{x}}
     h_2\,\tr \phi (\vec{x})^2
      + h_4 \left( \tr \phi(\vec{x})^2 \right)^2.
\ee

In the scaling limit (\ref {scaling}) explained in subsection 2.1, 
we drop those terms of the small $a$ expansion of $h_2/a^2$ and 
$h_4/a^2$ which vanish as $a\to 0$ and find
\bea
 h_4 &=& \frac {9} {16 \pi \beta _3^2}, \label {h4} \\
 h_2 &=& -\frac {9} {\pi \,L_0\,\beta _3} \biggl (c_0\log L_0+c_1
 \biggr ); \,\quad 
  c_0=1,\quad c_1=\frac {5}{2}\log 2-1.
 \label{h2} 
\eea 
We call these expressions the scaling parts of $h_2$ and $h_4$.
Let us sketch how these results have been derived and comment on them.\\

The coupling $h_4$ requires the 1-loop nonstatic contribution to the Green's 
function $\widetilde\Gamma_{ns}^{(4)} (\vec{p}=\vec{0})$ with four external 
$A_0$
fields at zero momentum. At $T=0$, the Feynman integrals involved in the
continuum  all are
of the type $\int d^3k/k^4$ and thus UV convergent but IR divergent. At $T\neq 
0$,
the restriction to non static modes ($k_0\neq 0$)
makes them IR finite. For these reasons, $h_4$ can be, and was, calculated 
directly in the continuum. Our result (\ref {h4}) coincides with that obtained 
by adapting to $D=3$ the calculation of Nadkarni \cite {nadkarni} 
for $D=4$. 

The case for $h_2$ is slightly more tricky. It is given by 
\be \label {h2pi}
h_2=-\frac {6} {\beta _3} \widetilde\Pi^{ns}_{00}(0),
\ee
where $\widetilde\Pi^{ns}_{00}(0)$ is the 1-loop contribution of the
non static modes to the vacuum polarization at zero momentum. 
Two Feynman graphs contribute, with one and two $A_i$ internal
propagators, respectively of the form 
$g_3^2\int^\Lambda d^3k/k^2$ and $g_3^2\int^\Lambda
d^3k\,k_0^2/(k^2)^2$, both linearly divergent
with the cut-off $\Lambda$. However, at $T=0$, gauge invariance guarantees
$\widetilde\Pi_{00}(0)=0$, so that
the two graphs cancel each other exactly there. At $T\neq 0$, the
cancellation is only partial, still leaving a 
logarithmic UV divergence. We thus computed $ \widetilde\Pi^{ns}_{00}(0)$
{\it on the lattice}, which provides a cut-off of order $\Lambda \simeq 1/a
=T\,L_0$ at high temperature according to Eq.(\ref {temp}), hence
the $\log (L_0)$ term in (\ref {h2}).

Applying the Feynman rules of \cite {frules}, the
finite lattice expression of $ \widetilde\Pi^{ns}_{00}(0)$
is found to be
\bea  \label{pi00_t}
    \widetilde\Pi^{ns}_{00}(0) = 
     \frac{3}{L_0\,L_s^2} \sum_{k_0\neq 0} \sum_{\vec{k}}
      \biggl\lbrace 
        \frac{2 c_{k_0}^2 \widehat{k}_0^2}
             {\left( \widehat{k}^2 \right)^2}
          - \frac{1 - \widehat{k}_0^2/2}
                 {\widehat{k}^2}
    \biggr\rbrace,  \\ \label{pi00}
    \widehat{k}_\mu =2\sin (k_\mu /2),\quad \widehat{k}^2=\sum_0^2
    \widehat{k_\mu}^2, \quad c_{k_\mu}= \cos (k_\mu /2).  \label{khat}
\eea    
The $L_s\to \infty$ limit is obtained by the replacement 
$(2\pi/L_s)^2\sum_{\vec{k}}\to \int_{-\pi} ^\pi d^2k$. After some
manipulations, one arrives at
\bea 
\widetilde\Pi^{ns}_{00}(0) &=& \frac {3} {L_0} \sum_{n_0=1}^{L_0-1}
\biggl (-\frac {1} {2} (1-2t)S(t)-t(1-t)\frac {dS(t)} {dt}\biggr ),\quad
t=sin^2(\frac {\pi \,n_0} {L_0}), \label {pist}\\
S(t)&\equiv&\frac {1}{4\pi^2} \int _{-\pi}^\pi \frac {dk1\, dk2} 
{2(1+t)-\cos (k_1)-cos(k_2)}. \label {st}
\eea
It is clear on the latter equation that $S(t)$ has a logarithmic
singularity at $t=0$.
The result (\ref {h2}) follows from (\ref {h2pi}) after the asymptotic 
behaviour at large $L_0$ and infinite $L_s$ of $ \widetilde\Pi^{ns}_{00}(0)$
is inserted.  Details on its derivation from (\ref {pist})
are given in the appendix. Here, we just  mention a useful
trick used to evaluate $\widetilde\Pi^{ns}_{00}(0)$  
, noting that Eq.(\ref {pist}) can be rewritten as
\bea \label {pix}
\widetilde\Pi^{ns}_{00}(0)&=&-\frac {3} {2\,L_0} \sum_{n_0=
1}^{L_0-1} \frac {d} {dx} G(x),\quad x=\frac {\pi\,n_0}{L_0}, \label{trick}
\\
G(x)&\equiv&\sin x \,\cos x \,S\bigr (\sin ^2x\bigl ).
\eea
At large $L_0$, 
$\sum_{n_0=1}^{L_0-1}\frac {d} {dx} G(x)$ is of order
$L_0/\pi \int _{\pi /L_0}^{\pi -\pi/L_0} d G(x)=
-2L_0/\pi G(\pi/L_0)$: The linear divergence in $L_0$ is cancelled
by the $\sin (\pi/L_0)$  factor in $G$ while the $\log L_0$  originates
from the logarithmic behaviour of $S(\sin^2 x)$ at 
$x\simeq \pi/L_0$.

\vspace{0.3cm}
\subsection {\bf  About the Reduced Action: Summary and Remarks} 

We first summarize what has been done above. The  QCD${}_3$ lattice action
is given by Eq.(\ref {wilson}). The parameters of physical relevance
are the lattice coupling $\beta _3$ and $L_0$. The observables we
consider are the Polyakov loops (\ref {ploop}) and their correlation
(\ref {plc}), and the spatial Wilson loops (\ref {wloop}).

Dimensional reduction leads to the 2D lattice effective action described
by Eqs. (\ref {seff}-\ref {h2}). It is a gauge invariant model
for a scalar $\phi$ in the SU(3) adjoint representation, whose ``Higgs
potential'' is truncated at order 4 in $\phi$ at high temperature and
low momenta. The lattice gauge coupling and scalar self couplings
$\beta _2,
h_2$ and $h_4$ are fixed by the 3D parameters Eqs. (\ref {beta2}, \ref
{h2}, \ref {h4}). The Higgs field is normalized in such a way its
kinetic term in the classical limit is $\tr(\partial _i\phi )^2$.

Let us now comment upon what we have obtained. First, as announced before, it
is consistent to truncate the effective action to order
$\phi ^4$, with $h_2$ and $h_4$ couplings computed at 1-loop order. 
To be more precise, power counting  and inspection show that 
$h_n/a^2$, with $h_n$ the coupling 
associated with any monomial of degree $n$ in $\phi$, has the 
following order of magnitude:
\bea
 n = 2 &:& \qquad  g_3^2 T \; + \; g_3^2\,T\, \log T\, 
     \biggl (1+\cO [(p/T)^2, g_3^2/T]\biggr), \\
 n = 4 &:& \qquad   g_3^4\, (1+\cO (g_3^2/T)), \\ 
 n \geq 6 \;\mbox{even} &:& \qquad 
           O\left( T^2\,(g_3^2/T)^{n/2}\right), \\
 n \geq 3 \;\mbox{odd} &:& \qquad 
           O\left( T^2\,(g_3^2/T)^{n/2} \, (p/T)\right).
\eea

The corrections for $n=2$ and 4 come from 2-loop order and non-local
effective interactions. It is thus consistent to 
 neglect the $n=6$ and higher power monomials, as well as all couplings 
 with $n$ odd, and to keep $h_2$ as
 given by Eq. (\ref{h2}). 

We next note that unlike $h_4$ which is
positive, insuring boundedness of the partition function at large
fields, the quadratic coupling $h_2$ is negative. The ``potential'' thus
have a shape typical for gauge symmetry breaking by Higgs mechanism.
Of course, this symmetry breaking is not expected to occur in 2 dimensions. 
In fact, the form (\ref {h2}) of this coupling can be seen as a
counterterm  for the logarithmic UV divergence of the 2D model, here prescribed 
in advance by the UV regularization of the 3D model, while the actual IR
behaviour of the theory is highly non perturbative.

In the static gauge chosen, the 3D Polyakov loop 
operator is given in 2 and 3 dimensions 
by the same function (\ref {statl}) of the $A_0$ field.
 According to the normalization (\ref {phi})
of the $\phi$ field and to Eqs. (\ref {scaling}, \ref {beta2}), it reads
\bea
L(\vec{x})&=&\frac {1}{3}\tr\exp( i\,\phi (\vec{x})\,/\sqrt\tau), \label
{plphi}\\
\tau &=&\frac {\beta _3}{6L_0}=\frac {\beta _2}{6L_0^2}
=\frac {T}{g_3^2}. \label {tau}
\eea
Along a line $\tau=const.$ (constant physics, see subsection 2.1), the operator 
used to probe dimensional reduction remains unchanged, and it is thus 
consistent to compare the Polyakov loop correlations measured in 3 and 2 
dimensions at the same fixed $\tau$ value. The continuum limit is  
$L_0\to \infty$, and constant (large distance) physics at large $L_0$ can 
be checked within either one of the two models.  

Let us then consider the effective action $S_{eff}^2$ in terms 
of the dimensionless temperature $\tau$, Eq. (\ref {tau}), and of the 
lattice parameter $L_0$.  After setting 
$a^2g_2^2=6/\beta _2$, and using Eqs. (\ref{beta2}, \ref {h2}, \ref
{h4}, \ref {tau}), the three dimensionless couplings $(ag_2)^2, h_2$ and
$h_4$ read 
\bea
(ag_2)^2&=&\frac {1}{\tau L_0^2} ,\\
h_2&=&-\frac {3}{2\pi } \frac {c_0\log L_0 +c_1}{\tau L_0^2}, \\
h_4&=&\frac {1}{64\pi} \frac {1}{\tau L_0^2} \frac {1}{\tau}.
\eea
We see that, except for the $\log L_0$ in $h_2$, 
these three couplings all scale as $1/L_0^2$ with $L_0$,
 while as a function of $\tau$, $h_4$ is  
 $\tau$ times smaller than the two other ones. 

Let us finally write down the effective
 Lagrangian $\cL _{eff}$, as obtained from the small $a$
  expansion of the effective action $S_{eff}^2$. In $S_{eff}^2$, we
 make the substitution $A_i\to ag_2A_i$, replace $\sum _{\vec {x}}$ by 
$a^{-2}\int d^2x$, use $T$ and $g_2^2$ as parameters instead of
 the lattice $\beta _3$ and $L_0$, and take the limit $a\to 0$
 (but in $\log L_0\equiv -\log aT$ in the $\phi ^2$ term). We obtain 
 \bea
 \cL _{eff}&=&\frac {1}{4} \sum _{c=1}^8 F_{ij}^c\,F_{ij}^c + \tr
[D_i\phi]^2 +\frac {g_2^2}{32\pi}\biggl (\frac {g_2}{T}\biggr )^2\,
\tr \,\phi ^4 +\cL_{CT}, \\
 D_i \phi &=& \partial_i \phi + ig_2 [A_i,\phi],
\nonumber \\
 F_{ij} &=& \partial_i A_j - \partial_j A_i + i g_2 [A_i, A_j], \nonumber \\
 \cL _{CT} &=& -\frac {3g_2^2}{2\pi}
 \biggl [-\log (aT)+5/2\log 2-1\biggr ]\,\tr \,\phi ^2. \label{counter}
 \eea

This is a 2D, SU(3) gauge invariant Lagrangian for an adjoint scalar
$\phi$, but it is far from being the most general one. The gauge coupling 
$g_2$, with its canonical dimension one in energy, sets
the scale. The non kinetic quadratic term is 
the counterterm $\cL_{CT}$, suited to a lattice UV regularization with spacing 
$a$. The quartic self
interaction $\lambda _4=g_2^2\times (g_2/T)^2$, generically
a free parameter, here goes to zero as $T\to \infty$ in units of $g_2$,
and presumably plays a marginal role.
We also note the absence in the model of odd powers $\phi ^{2k+1}$, 
allowed by gauge symmetry for $k>0$. The corresponding $Z_2$ symmetry might 
however be spontaneously broken  in some subspace of the unrestricted parameter
space  {$g_2, h_2, h_4$}. The similar problem in $4\to 3$ QCD reduction 
has been studied in Ref. \cite {kajantie2}. 
For the case of SU(2) in (3+1)D, $Z_2$ is
the center of the gauge group so that $Z_2$ breaking is also gauge symmetry
breaking, a subject previously discussed in Refs. \cite {kark}, \cite
{thomas2} and \cite{kajantie1}.

\section{\bf The Numerical Simulations in (2+1) and 2 Dimensions}
\vspace{0.5cm}

We simulated the reduced theory using the same method as in 
reference \cite{su3}. For updating the gauge fields $U$ we employed a
mixed algorithm, where new trial SU(3) matrices were generated by the
heath--bath algorithm and are accepted according to a Metropolis
condition on the hopping term. We used the standard Cabibbo-Marinari
pseudo heath--bath algorithm based on updating the SU(2) subgroups of
a SU(3) matrix \cite{heathbath}. The new SU(2) sub--matrices were
generated using the Kennedy--Pendleton algorithm \cite{kp}.  After each
subgroup update the resulting matrix was subjected to the Metropolis
question on the hopping term.  This approach results in an acceptance
rate close to 95\%.

Alternatively we tried the multi--hit Metropolis algorithm. An update
of a single link variable was attempted several times (typically
eight) before attempting to update next variable. The trial matrices
were generated according to~: $U\rightarrow \exp(i A) U$ where $A$ was
an hermitian matrix generated from the distribution $\exp(-\epsilon\;
\tr A^2)$. The parameter $\epsilon$ was tuned to obtain the acceptance
ratio around 50\%. Because generation of trial matrices is expensive
in CPU time (matrix exponentiation) we took the values from a table
that was regenerated every five sweeps. The autocorrelation times for
this algorithm were comparable to the heath--bath but the CPU time
required to perform a metropolis update was approximately two times
bigger.

The scalar fields were updated using the multi--hit 
metropolis. At every point we tried typically eight updates. New scalar fields
were obtained from $A\rightarrow A+\delta A$ where $\delta A$ was 
generated according to $\exp(-\epsilon\; \tr A^2)$.  For each set of
coupling constants, $\epsilon$ was tuned to obtain the acceptance
ratio around 50\%. One sweep over the lattice consisted of one update of 
gauge fields followed by one update of scalar fields. After each sweep we 
measured ``control'' variables~: expectation value of plaquette, $\tr A^2$,
$\tr A^3$ and $\tr A^4$.  

All runs except for scaling test (see the next section) were
performed on $32\times32$ lattices with the parameter $L_0$ set equal to
4.  The dynamics of the gauge and Higgs sectors turned out to be quite
different.  The integrated autocorrelation time of the plaquette operator
was typically non--measurable ($\lesssim1$) while for $\tr A^2$
it varied
between 100 and 400 sweeps when $\beta_3$  was varied from $21.0$ to
$346.0$.  For each value of $\beta$ we have collected typically $10^6$
sweeps.

In three dimensions we used the data previously gathered on the
Quadrics APE supercomputer \cite{lego} on $32\times 32\times 4$
lattices. 

The main quantities measured were the average $\av{L(\vec{x})}$ of the Polyakov
loops, their on-axis 2-body correlations $P(r)$, and the average
$\av{W(R_1,R_2)}$ of spatial Wilson loops. 
In three dimensions they are  defined by (\ref{wloop}), (\ref{ploop}) and
(\ref{plc}).  The only difference in two dimension is that the Polyakov loop
is given by (\ref{statl}). 
When necessary, an index $D=2+1$ or $2$ specifies the dimension
considered.  Both 2D and 3D data were analyzed in the same way as
to minimize the systematic errors.

In 2D simulations Wilson loops were measured every ten sweeps and
blocked by 50. Integral autocorrelation time for the measurements of
$14\times14$ Wilson loops was, in the worst case, of the order of one
and was negligible in the majority of cases. The measurement routine
is quite expensive and consumed almost half of the CPU time. 

In order to extract the string tension, local potentials were first extracted 
from the ratios~:
\begin{equation}
V(R,R')=\log\frac{\av{W\left(  R,R'\right)}  }{\av{W\left(R,R'+1\right)}  }.%
\end{equation}
By definition, the potential is
\begin{equation}
V(R)=\lim\limits_{R'\rightarrow\infty}V(R,R').
\end{equation}
For each given $R$, $V(R,R')$ was found to decay exponentially to
a constant in $R'$. In practice, the constant regime
is reached within errors above $R'=3$ and the potential was
fitted in the range $R'\in\left[ 4,12\right]$. 
The string tension $\sigma$ was then obtained from the ansatz%
\begin{equation}\label{vfit}
V\left( R\right) =V_{0}+\sigma R.
\end{equation}

A fit to the above linear potential is sufficient to obtain a stable
value of $\sigma$ within errors when it is performed in the range from
$R_{min}=2$ to $R_{max}$.  The $R_{max}$ values, depending upon the
value of $\beta _3$, varied from $10$ to $14$.  The $\chi^2$ value in
the 3D case was of the order of one, and about ten times less in the
2D case, likely due to large correlation between points.  

The errors on $\sigma$ were calculated by the triple application of
the jack-knife method. In order to calculate the jack--knife estimate
of the error on $\sigma$ from the fit to the formula (\ref{vfit}) we
used 16 jack--knife copies of the potential $V(r)$ with errors.  To
estimate the errors of every copy we applied the jack--knife algorithm
with the same block size again to each of the corresponding 16 data
sets, thus generating 15 copies of every set.  However, in order to
calculate the $V(r)$ we needed the ratios $V(R,R')$ again with
errors. Calculation of these errors was done by a third and last
jack--knife step applied to each of $16\times 15$ data sets obtained in
the previous step.

Polyakov loops correlators were also measured every ten
sweeps. While, due to the large auto--correlation time, this was 
in a sense an ``oversampling'', it did not take much of the CPU
time. The measurements were blocked by 50 and then written
out. Afterwards the data were again blocked into 25 blocks.  These 25
blocks were used to calculate the connected correlators with errors
estimated by jack-knife algorithm. We also tried different numbers of
blocks and did not observe any significant change in the results.  The
double jack-knife algorithm, similar to the one described for the
Spatial Wilson loops, was used to estimate the error for fits of the
connected correlator to the formulae (\ref{pole}, \ref{cut}) used to
extract a screening length (see next section). The fits were done in the 
interval $r\in [r_{min}=4,15]$.  We have checked that varying $r_{min}$ 
from 3 to 6 does not change the results.

\section{\bf Results and Discussion}

We now present our results, discussing successively
dimensional reduction (how well does the effective 2D model describe
quantitatively the large distance physics of high temperature 
QCD${}_3$), scaling (is the parameter region explored close to the
continuum limit), the screening length in the Polyakov loop channel 
and the spatial string tension.

\subsection{Dimensional Reduction}
Let us compare the Polyakov loop correlations $P_{2+1}(r)$ extracted
from Ref. \cite{lego} and $P_2(r)$ measured in our simulation. A
conservative statement on dimensional reduction is to say that the
lowest physical state coupled to $L(\vec{x})$ is the same in both
cases, so that the two functions must have the same $\it shape$ in $r$
at large $r$.  A stronger statement is that, to the extent that the
weight associated with $S_{eff}^2$ is a good approximation to the
integral over the non-static fields of the $(2+1)D$ weight, in the
small $p/T$ regime at least, the averages with respect to the two
weights of a static operator such as $\av{L(\vec{x})\,L(\vec{y})}$ are
$\it equal$ at large $\vert \vec {x} -\vec {y}\vert$. We will show
that this latter situation is indeed approached at large enough
temperature.

We recall that temperature is given in units of the gauge coupling by
\be
\tau \equiv \frac {T}{g_3^2}=\frac {\beta _3}{6L_0}.
\ee
Using $L_0=4$ fixed, we investigate the temperature dependence of the
correlations by varying $\beta _3$ in the range  $[21, 173]$. 
High temperature means $\beta _3$ sufficiently 
larger than the transition point in $(2+1)D$, which was found to occur at 
$\beta _c=14.73$ for $L_0=4$ \cite {lego}. 
\begin{figure}
\begin{center}
\psfrag{xlabel}[cb][cb][1][0]{$r$}
\psfrag{ylabel}[c][t][1][-90]{}
\psfrag{s0label}[r][r][.8][0]{(2+1)D}
\psfrag{s1label}[r][r][.8][0]{2D}
\includegraphics[clip,width=14.5cm,bb=28 35 540 308]{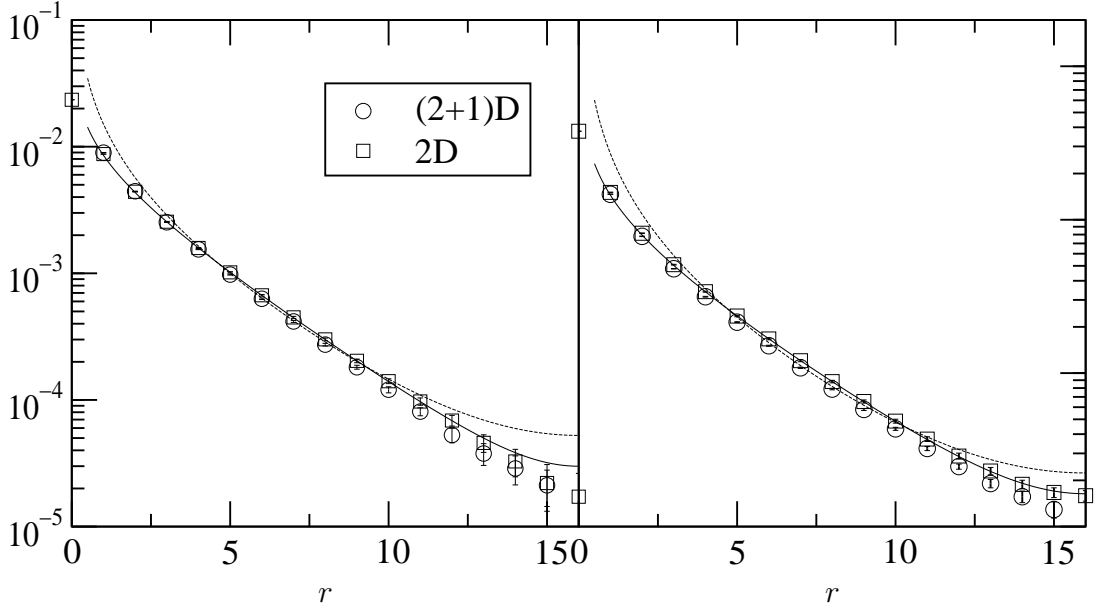}	
\end{center}
\caption{Polyakov Loop Correlations $P_D(r)$, $D=2+1$ (circles) and
$2$ (squares), for $\beta_3$ equal to $29$ (left) and $84$ (right).
Distance $r$ is given in lattice units.  In both cases $L_s=32$ and
$L_0=4$.  The $2D$ data were produced using $h_2$ from Eq.
(\ref{h2}).  The continuous and the dashed lines result 
respectively from fitting formulae (\ref{pole}) and (\ref{cut}) to the $2D$ 
data.}
\end{figure}

Fig.1 compares the correlations $P_{2+1}(r)$ and $P_2(r)$ for
$\beta_3$= $29$ and $84$. The curves are fits of the
$P_2(r)$ function, to be explained below.  We see that
the two correlations are extremely close one to the other: not only they
have the same {\it shape} at large $r$, which is the primary prediction of
dimensional reduction, but also nearly the same {\it normalization}.
As announced, this finding favours dimensional
reduction in the strong sense. In addition, we notice that this
agreement between the two ways of computing the Polyakov loop
correlation extends down to fairly small values of $r$.  Since the
normalization is set by $|\av{L(\vec{x})}|^2$, it shows that,
although it is local, the Polyakov operator is not very sensitive to
the short wave length terms omitted in the effective action. That it
is the case is born out by the data of Table 1, where our results for
$\av{L_2(\vec{x})}$ and $\av{L_{2+1}(\vec{x})}$ are presented. Relatively to
their distance to one (their common value at $\beta _3=\infty$), their
difference, already $\sim 6\%$ at $\beta _3$=21, decreases to $\sim 0.6\%$
at $\beta _3=$173.
\begin{table}[h]
\begin{center}	
\begin{tabular}{||l||l||l||}\hline\hline
$\vphantom{\bigg(}\beta_3$ &\multicolumn{1}{c||}{$|\av{L_2(\vec{x})}|$} 
&\multicolumn{1}{c||}{$|\av{L_{2+1}(\vec{x})}|$}\\\hline\hline 
21.0 &	0.56002 (62)& 0.53467 (16) \\\hline 
29.0 &	0.67007 (25)& 0.66120 (13) \\\hline 
42.0 &	0.76397 (20)& 0.76130 (12)  \\\hline 
84.0 &	0.87392 (11)& 0.87435 (6) \\\hline 
173.0&	0.93494 (13)& 0.93530 (11)\\\hline \hline
\end{tabular}
\end{center}
\caption{The average Polyakov loop as a function of
$\beta_3$ in (2+1)D and 2D.}
\end{table}
\begin{figure}
\begin{center}
\psfrag{xlabel}[b][b][1][0]{$r$}
\psfrag{ylabel}[c][c][1][0]{}
\includegraphics[width=12cm,clip,bb=32 70 537 507]{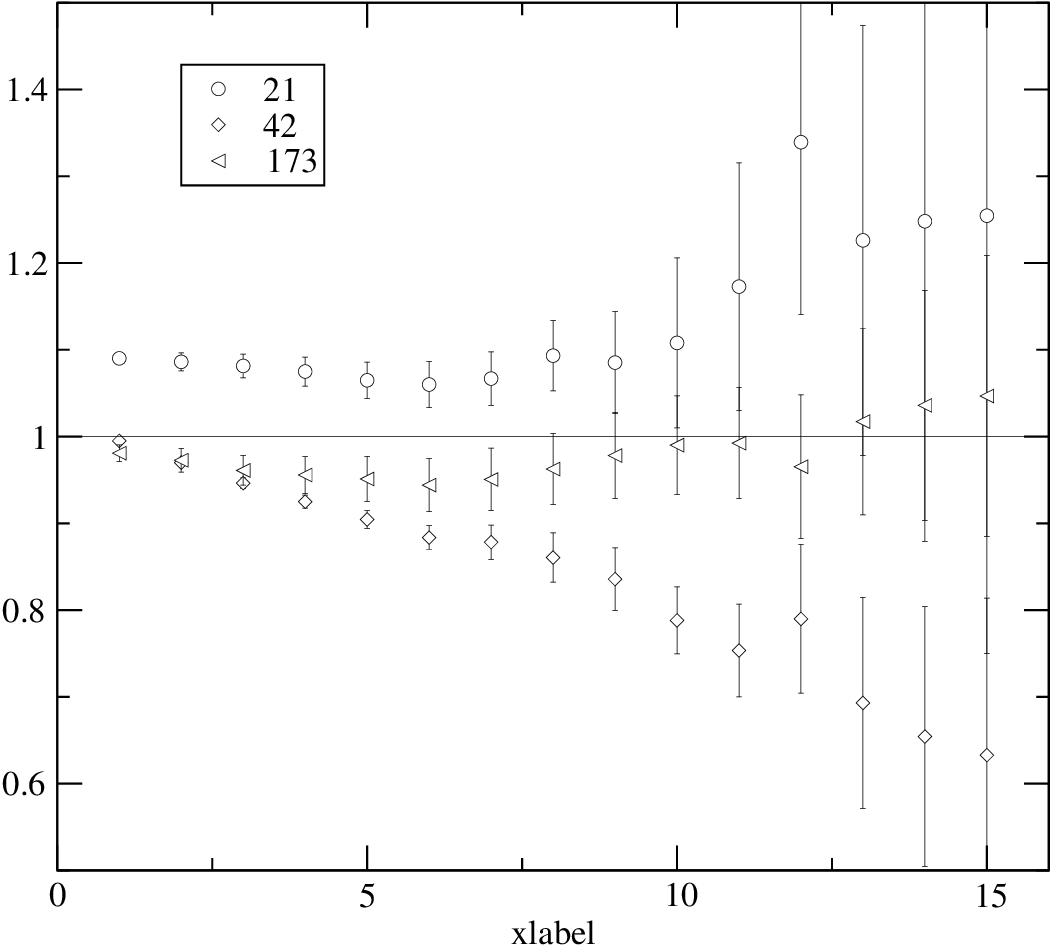}%
\end{center}
\caption{Ratios $P_{2+1}(r)/P_2(r)$ of Polyakov Loop correlations 
as functions of the lattice distance $r$ for $\beta_3=21, 42, 173$.}
\end{figure}

In Fig. 2, we illustrate more in details, 
for the $\beta _3$ values not reported 
on in Fig. 1, how well the $2D$ and $(2+1)D$ correlations compare. 
Their ratio $P_{2+1}/P_2$ is plotted against the distance $r$ in lattice
units. These data definitely support the statement that $P_{2+1}/P_2$
remains quite flat and close to one in the whole $r$ and $T$ range. 
Recalling that the lowest $T$ value is only $\sim 1.5$ times $T_c$, 
and that the correlation functions decrease in $r$ by about three 
orders of magnitude, we conclude that the effective local 2D action 
reproduces the $(2+1)D$ Polyakov loop correlations with a 
remarkable accuracy, soon above the transition, and down to distances 
even shorter than $1/T$.

\subsection{Scaling}
Given the temperature, the continuum limit is approached by taking 
\be
\tau =\frac{\beta _3}{L_0}=\frac{T}{g_3^2} \quad\mbox{fixed}, \qquad \qquad
L_0=\frac {
1}{aT} \quad\mbox{large}.
\ee
\begin{figure}
\begin{center}
\psfrag{ylabel}[c][c][1][0]{}
\psfrag{xlabel}[c][c][1][0]{$r$}
\psfrag{biggerandbigger}[l][l][.8][0]{$\beta_3=58,\;L_0=8$}
\psfrag{smaller}[l][l][.8][0]{$\beta_3=29,\;L_0=4$}
\includegraphics[width=12cm]{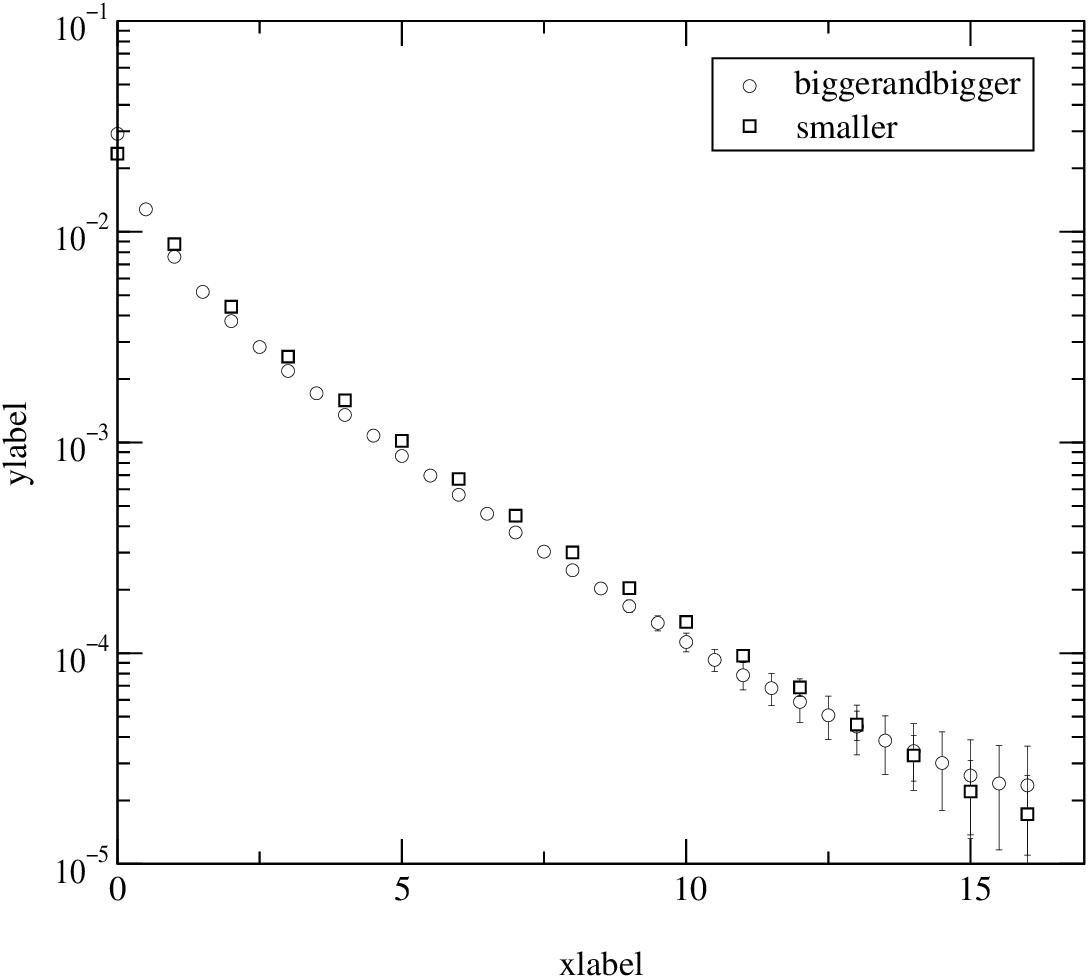}%
\end{center}
\caption{Comparison of the Polyakov Loops Correlations in the 2D model
for the two sets of parameters $[\beta_3,L_0,L_s]=[29,4,32]$ and
$[58,8,64]$ i.e. for constant $\tau$ (\ref{scaling}). The values of $r$ for
$L_0=8$ (circles) are scaled down by a factor two in order to maintain the
same physical scale.\label{fscaling}}
\end{figure}

Scaling is thus the statement that, at fixed $\tau$, the physics do
not change with $L_0$, if $L_0$ is large enough. We checked that by
comparing the Polyakov loop correlations in $2D$ for two sets of lattice
parameters, namely $[\beta _3, L_0,L_s]=[29,4,32]$ and
$[58,8,64]$. Given $T$, doubling $L_0$ means dividing the lattice
spacing $a$ by two, so that the ${\it physical}$ spatial size $aL_s$
of the lattice is unchanged. The resulting correlations are presented
in Fig.~\ref{fscaling}\, showing a very similar shape as a function of
the physical distance. It is again instructive to consider their ratio,
found to be quite flat at all distances (Fig.~\ref{fratio}): Within
errors, there is no sizable deviation from scaling. Note that this
constant ratio is not one. It is given by that of the corresponding
values of $|\av{L(\vec{x})}|^2$, which is not a physical quantity
(remember 
that the $2D$ effective Lagrangian contains an explicit (logarithmic)
dependence on $a$ via its counterterm Eq. (\ref{counter})).
\begin{figure}[!t]
\begin{center}
\psfrag{xlabel}[c][c][1][0]{$r$}
\psfrag{ylabel}[c][c][1][0]{}
\includegraphics[width=12cm,bb=15 80 537 495]{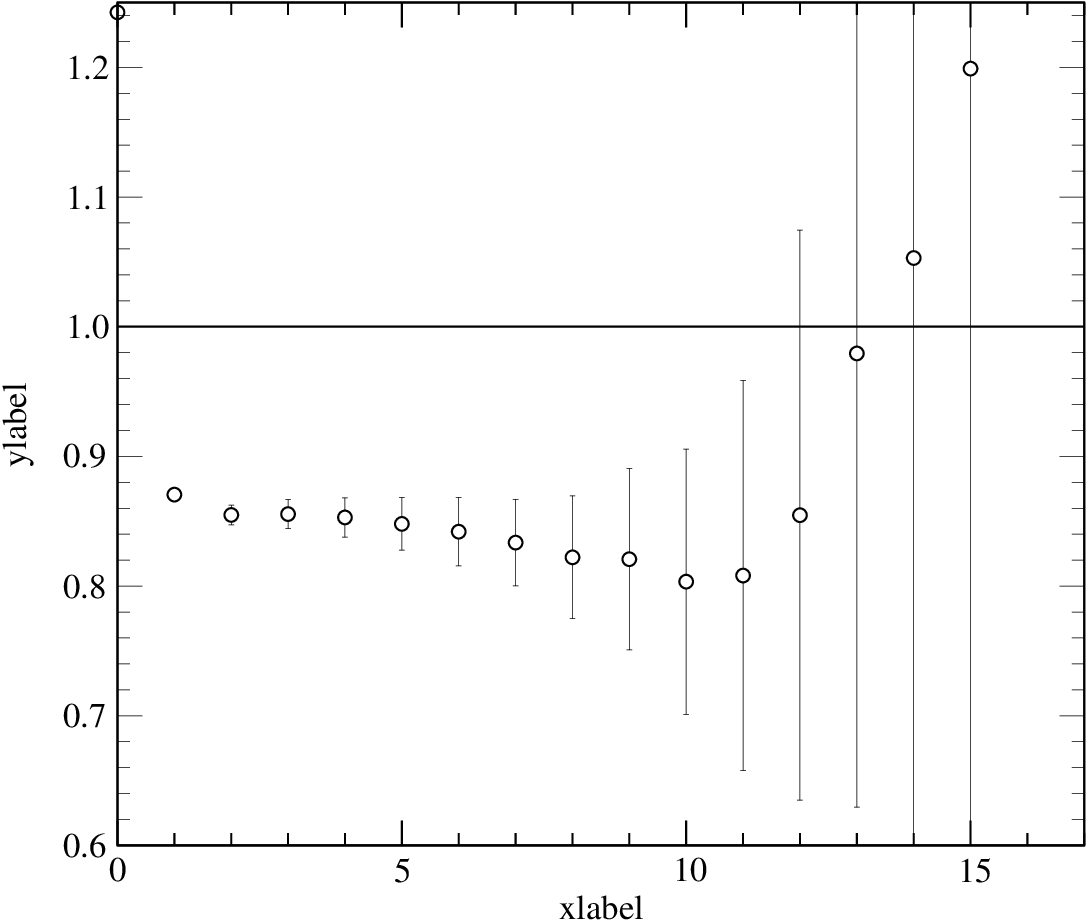}%
\end{center}
\caption{The ratio between the two Polyakov loop correlation functions
presented in figure~\ref{fscaling}.\label{fratio}}
\end{figure}

Hence scaling is verified in the range of interest. It justifies
keeping $L_0=4$, which is less expensive in computer time and
allowed us to use the existing data of \cite{lego} in $(2+1)D$.

\subsection{Screening Lengths}

The fast decay with $r$ of the correlation functions is consistent
with the existence of a finite spatial correlation length, $\xi _S$, 
associated with the quantum numbers of the Polyakov loop operator, 
a colour singlet scalar. A mass can be defined as usual by $M_S=\xi _S^{-1}$.
In the similar situation with one more dimension, perturbation theory
is invoked to state that $M_S$ is $2M_E$, twice the so called
electric screening mass, because the lowest ``state'' coupled to the
loop is a two electric gluon state. If this is so, the $P(r)$ correlation 
is expected to be proportional at large r to the square of the correlation 
function for a one particle state of mass $M_E$. In the present case
however, the infrared sector is more complicated and perturbation theory
is doubtful. In particular we let open the possibility that there
exists in the PL channel an independent screening mass $M_S$,
associated with a true thermal excitation of the $2D$ system. 
In the latter case, we parametrize the data according to 
\be \label{pole}
P_D^{(m_S)}(r)\simeq
c\,\Biggl (\frac{1}{[m_Sr]^{1/2}}e^{- m_S\,r}+
\frac{1}{[m_S(L_s-r)]^{1/2}}e^{- m_S (L_s-r)}\Biggr ), \label{vrsym}
\ee
while in the former case one rather expects
\be \label {cut}
P_D^{(2m_E)}(r)\simeq
c'\,\Biggl ( \frac{1}{[m_Er]^{1/2}}e^{- m_E\,r}+
\frac{1}{[m_E(L_s-r)]^{1/2}}e^{- m_E(L_s-r)}\Biggr )^2. \label{cutsym}
\ee
The $m$ symbols  denote masses in lattice units, i.e. $m\equiv aM$. These
parametrizations respect the lattice symmetry $r\to L_s-r$. Since we 
have no direct access to $m_E$, the two above expressions differ 
in shape through the prefactors, $r^{-1/2}$ and $r^{-1}$ respectively.
We checked that for $r>1$, lattice
artefacts are negligible in the mass range considered. This was achieved
by comparing Eq. (\ref{pole}) to the lattice propagator 
\bea
P_{Latt}(m_S,r)=\frac {1}{L_s^2} \sum _{p_1,p_2} \cos(p_1r) 
\widetilde P_{Latt}(m_S,\vec {p}), \\
\widetilde P_{Latt}^{-1}(m_S,\vec {p})=\widehat {p}^2+4\sinh (m _S^2/4).
\eea
in the notations of Eq. (\ref {khat}).  
We carefully
analyzed our numerical data, and {\it we find that the ansatz
(\ref{pole}) is
by far the best one}, giving a good fit of the data down to small $r$
values. For illustration, the continuous curves of Fig. 1 are fits of the
form (\ref {pole}) to $P_2(r),r\geq r_{min}$, with $r_{min}=4$, and
changing $r_{min}$ from 3 to 6 does not change the answer for $m_S$
within errors. The only deviation which we observe 
is at (very) short distances, where some more massive
contribution may be present, if not an artefact of the
lattice UV regularization. It is anyway too small to be reliably 
analyzed. On the contrary, fits to the same data of the form Eq.
(\ref {cut}) lead in Fig. 1 to the dashed curves, which are clearly 
not acceptable.
\begin{figure}
\begin{center}
\psfrag{xlabel}[c][c][1][0]{$g^2_3/T$}
\psfrag{ylabel}[c][c][1][0]{$M_S/\sqrt{Tg^2_3}$}
\includegraphics[width=12cm]{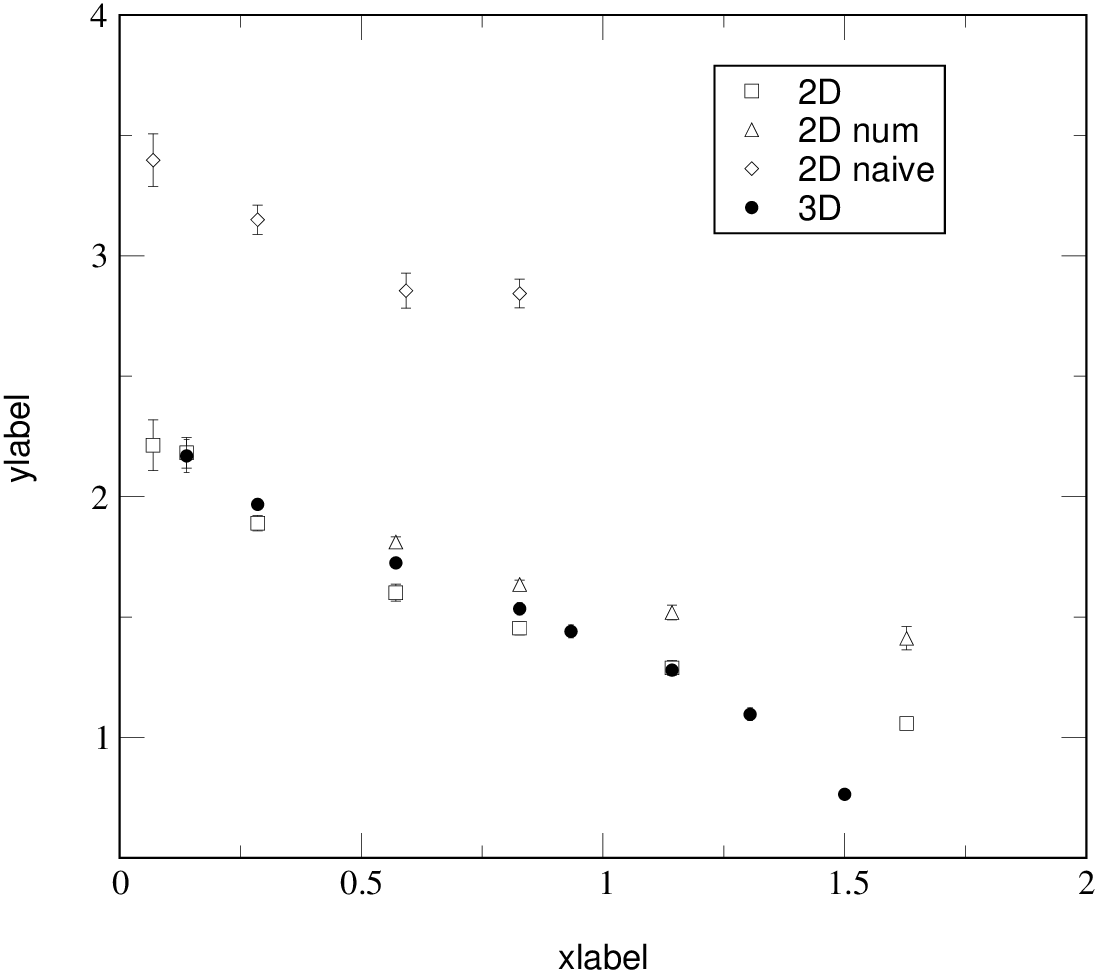}%
\end{center}
\caption{Physical screening masses $M_S$ in units of $g_3\sqrt T$
versus $g_3^2/T$, in $(2+1)D$ (black points) and $2D$ (squares). 
Also shown are the masses obtained 
with the numerical value of $h_2$ from Eq. (\ref{h2pi})
(triangles) instead of its asymptotic expression (\ref{h2}), and with 
the ``naive'' reduced action: no Higgs
potential, $h_2=h_4=0$.}
\end{figure}
The results of systematic fits of expression (\ref {pole}) to all
available correlations $P_2(r)$ and $P_{(2+1)}(r)$ are presented in Fig.5. 
The corresponding dimensionless quantities	
$M_S/(g_3\sqrt T)\equiv m_S\sqrt {L_0\beta _3/6}$ are plotted 
versus $g_3^2/T\equiv 6L_0/\beta _3$. Also shown for comparison are the 
masses resulting from a few
simulations made with the so-called naive effective action
($h_2=h_4=0$). The latter are far away from the former, showing the 
important effect of taking into account the non static degrees 
of freedom. We also made runs where, instead of using the so-called
scaling form (\ref{h2}) of $h_2$, we took the value 
obtained numerically from Eqs. (\ref{h2pi}, \ref{pi00}). The
results are similar, especially at high temperature, but anyway 
distinguishable within our statistical accuracy. This shows a
great sensitivity of the static properties to the quadratic counterterm
(\ref{counter}), an interesting feature per se.

The scaling properties observed in subsection 4.2 of course reflect
themselves in the mass values. For the two cases compared there, we
find $M_Sa$=0.331(6) and 0.169(3): the lattice spacing is reduced 
exactly by two within errors.

Of course the fact that dimensional reduction works well for masses  
directly follows from its success for correlations. We preferred to
illustrate it first on the latters, as done in Fig.2, that is independently
of any interpretation of the nature of the observed screening lengths.
What we now get in addition is that the main signal observed in the
Polyakov loop channel favours the existence of a true colour singlet
excitation of the high temperature gluon system, rather than it
reflects the existence of an electric screening length (Debye screening)
in the one longitudinal gluon channel.

\subsection{The Spatial Wilson Loop}
\begin{figure}[!t]
\begin{center}
\psfrag{ylabel}[c][c][1][0]{$\sqrt{{\sigma}/{Tg_{3}^{2} }}$}
\psfrag{xlabel}[c][c][1][0]{${g_{3}^{2}}/{T}$}
\includegraphics[width=12cm]{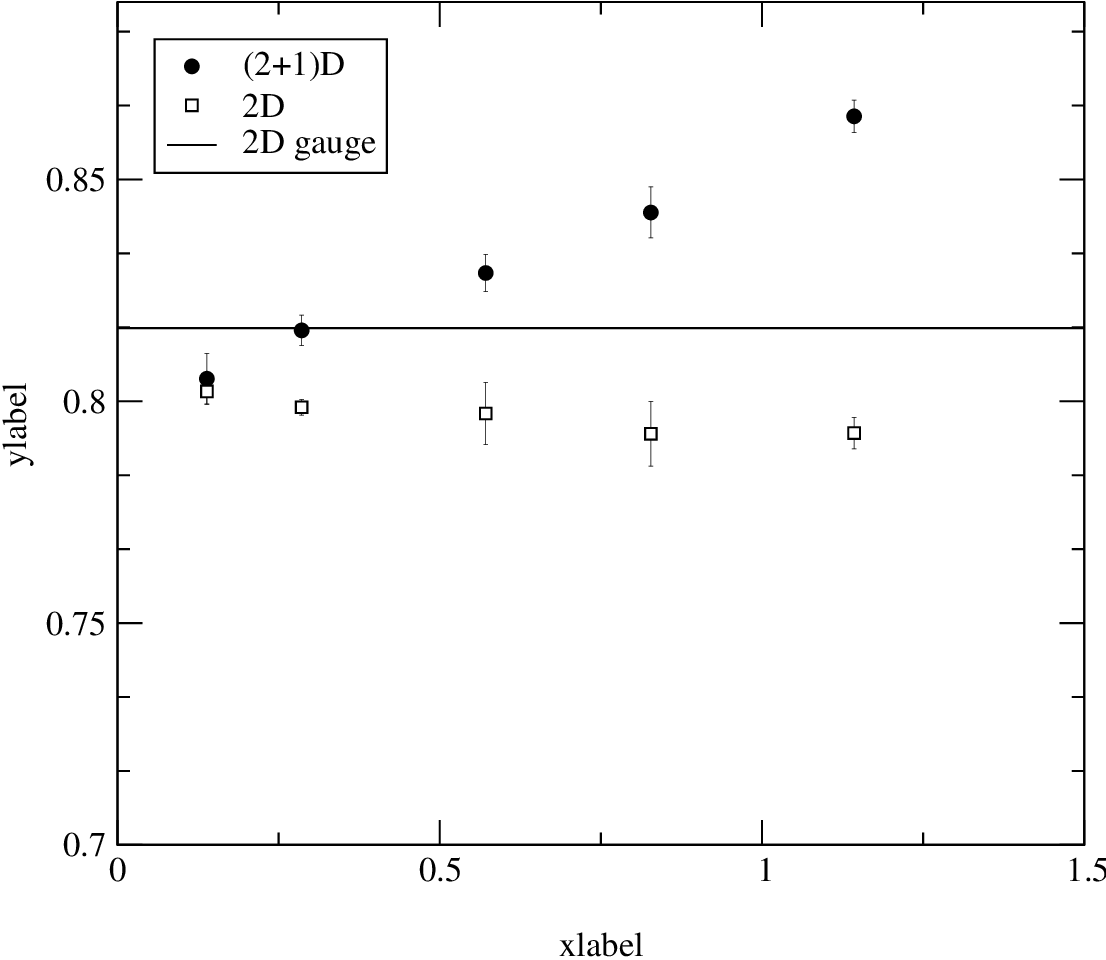}
\end{center}
\caption{The square root of the physical  string tension in units of 
$g_3\sqrt{T}$ as a function of $g_3^2/T$ in (2+1)D (filled circles) 
and 2D (squares). The line denotes the scaling limit $\sqrt {2/3}$
in 2D pure gauge theory (\ref{sigma02}). }
\end{figure}
The spatial Wilson loop $W$ in 3D is not a static operator,
and thus not predicted to assume the same value at high
temperature  and in the 2D reduced model. However i) a non zero spatial 
string tension $\sigma _{2+1}$ is known to exist \cite{lego} above 
the deconfinement temperature, ii) the pure gauge 2D theory is 
confining and produces a string tension $\sigma _{2}^0$ \cite{gross},
and iii) in two dimensions we expect confinement and thus a finite string 
tension $\sigma _{2}^{\phi}$ to survive when the Higgs field $\phi$ is 
turned on.

It is thus interesting to compare these three quantities as a function
of the temperature of the $(2+1)D$ model. For this comparison, we take
$a^2\sigma _{2+1}$ and $a^2\sigma _{2}^{\phi}$ from the analysis of  
$W$ data provided by \cite{lego} and by our simulation, whereas
$a^2\sigma _{2}^0$ is computed analytically at large $\beta _2=L_0\beta _3$.
This is obtained by expanding in $1/\beta _2$ the one plaquette SU(3) 
partition function (Eq. (\ref{sw2}) for one single site).
The SU(3) integral is performed following \cite {lang}, and one finds
\be \label {sigma0}
a^2\sigma _{2}^0=\frac {4}{\beta _2}+\frac {7}{\beta _2^2}+\cO (\beta _2^{-3}).
\ee
The next terms are easy to derive, but not required for our purpose.
This expression can be equivalently rewritten
\be\label{sigma02}
\frac {\sigma _2^0}{g_3^2T}=\frac {2}{3}+\frac {7}{36} \frac
{g_3^2}{T}(aT)^2+\cO \Biggl [(\frac {g_3^2}{T})^2(aT)^4 \Biggr ],
\ee
showing that $\sigma _2^0$ scales as $2g_3^2T/3$, 
up to scaling violations at finite $T$ of order 
$(aT)^2=1/L_0^2$. The quantities reported in Fig. 6 versus $g_3^2/T$  
are the
values of $\sqrt \sigma$ in units of $g_3\sqrt T$, which we thus
compare to $\sqrt {2/3}$. 
The numerical values of ${\sigma _2^{\phi}/(g_3^2T)}$ and 
$\sigma _{2+1}/(g_3^2T)$ were obtained from the simulations
for $L_0$=4. From the 2D point
of view, we see that introducing the $\phi$ field modifies the picture
quite weakly: no sizable $g_3^2/T$ dependence around a value close to
$\sqrt {2/3}$. The behaviour for $\sigma _{2+1}$ is substantially
different, with a sizable slope in $g_3^2/T$. The difference 
observed between $\sigma _{2+1}$ and $\sigma _2^{\phi}$ may be
understood as a consequence of $W$ not being a static operator:
the average of its non static modes with the $(2+1)D$ weight are missing
in its calculation with the effective action. However, scaling
violations may also contribute differently to these two $\sigma$'s,
so that definite conclusions require complementary simulations.
Let us finally recall that systematic errors can be present in $\sigma$
measurements, so that we cannot infer from the data presented
in Fig. 6 whether or not their infinite $T$ limit is the same, and equal
to $2/3$.

\section {\bf Conclusion}
\vspace{0.5cm}

In this article we have described a detailed investigation of the validity
of dimensional reduction for 
SU(3) gauge theory in 2+1 dimensions at high temperature.
We have constructed the reduced model for the static variables
in the same way as it was previously done in \cite{su2,su3,qcd}
on the reduction of QCD in 3+1 dimensions. 
This means that the integration over the
non static modes has been performed perturbatively, and the
effective couplings have been kept to one loop order for the
two point and four point functions. Higher order contributions
 go to zero at high temperature. Higher derivative couplings
are unimportant at large distances.

We investigated the validity of this approximation by calculating
the correlation function between Polyakov loops, and the
spatial string tension. For the 2+1 dimensional theory and
the corresponding 2 dimensional reduced adjoint Higgs-gauge
model we were able to obtain very precise numerical results.
Therefore we could make a detailed comparison, including
a scaling analysis which strongly supports the assumption that
our lattices correspond to a large enough value of the temporal extent
$L_0$.

We have found that for the correlation between Polyakov loops
the dimensional reduction works very well down to temperatures
$T \approx 1.5 T_c$, even at distances down to or below $1/T$. 
And it does even better than in the 3+1 dimensional case.

The correlation between Polyakov loops is well described by
a simple pole in momentum space, even at high temperatures.
This is in some contrast to 3+1 dimensional QCD, where at high
temperatures the data seem to indicate the presence of a cut,
corresponding to the exchange of two Debye screened gluons
\cite{gao,petereisz,frules,kacz}.
The difference may be related to the stronger infrared divergences
in 2+1 dimensions, making the simple perturbatively resummed
Debye screening picture invalid. A further investigation of this
behaviour in 3+1 dimensions, as well as the investigation of
other correlation functions in the lower dimensional case 
would certainly be very interesting.

The spatial Wilson loops in 2+1 dimensions do not correspond 
to a static operator. However, for the string tension, which is
extracted from Wilson loops of extent larger than $1/T$,
one can hope that the non static corrections are small. That
this is in fact the case is supported by the actual comparison
with the two dimensional model. For this operator, the Higgs
sector seems to have little influence, and there is a fairly good
agreement also with the pure two dimensional gauge theory,
which is analytically solvable, and where confinement
is given by the two dimensional Coulomb potential. The differences
between the string tension in the three cases considered is a few
percent. To judge if these differences are real continuum effects,
one must further study finite size and scaling corrections.

Our work shows that it may be possible to explain the
non perturbative features of 2+1 dimensional QCD 
in the deconfined phase with a relatively
simple two dimensional model.

\vspace{0.8cm} 
{\bf Acknowledgments} 
The authors thank Christian Legeland for the use of
his 3D SU(3) data and are grateful to Jean-Paul Blaizot for his
interest and clarifying comments. P.B. was supported by the Alexander von Humboldt
Foundation and partially by the KBN grant 2 P03B 019 17, and T.R. by a
Heisenberg fellowship.  K.P. was supported by DAAD. We also thank the
DFG for support under the contract Ka 1198/4-1. This project was
carried out in part on the new cluster computer ALiCE at COMPASS,
Wuppertal University, during its initial test phase. We thank the
staff of the ALiCE cluster-computer at Compass, University of
Wuppertal, for their support.

\appendix 
\section{Calculation of the $\phi ^2$ coupling in the Effective 2D Action}
\vspace {0.5cm}
We derive the result for $h_2$ announced in Eq.(\ref {h2}). According to
Eqs.(\ref {h2pi}, \ref{pix}, \ref{st}), the sum we need to perform in the large
$L_0$ limit is 
\bea
W\equiv -\frac {L_0} {3} \widetilde\Pi^{ns}_{00}(0) &=&\frac 
{1}{2}\sum_{n_0=1}^{L_0-1} \frac {dG(x)} {dx} ,\quad
x=\frac {\pi \,n_0} {L_0}, \label {appix}\\
G(x)&=&\sin(x) \cos(x)\, S(\sin^2 (x)) \label {fx}\\
S(t)&\equiv&\frac {1}{4\pi^2} \int _{-\pi}^\pi \frac {dk1\, dk2}
{2(1+t)-\cos (k_1)-cos(k_2)}. \label {apst}
\eea
In the latter integral, we replace the inverse denominator $D^{-1}$ by $\int 
_0^\infty dy\exp (-y\, D)$, integrate over $k_1,k_2$ and use the definition 
of the modified Bessel function $I_0$ to get
\be
S(t)=\int _0^\infty dy\, \exp [-2(1+t)y]\, I_0^2(y). \label {bessel}
\ee
A closed form for this integral is (see for example Eqs.(6.612.4) in
 \cite {grad}
and (17.3.9) in \cite {abram})
\be
S(t)=\frac {(-z)^{1/2}}{2} F(1/2,1/2;1;z),\quad 
z=-\frac {1}{t(t+2)}\label {hyper} 
\ee
which relates the behaviour of $S(t)$ at low $t$ to that of the
hypergeometric function $F(1/2,1/2;1;z)$ at large negative $z$. One finds
(Eq.(15.3.13) of Ref.\cite{abram})
\be
S(t)=\frac {1}{2\pi ^2}\sum_{n=0}^\infty 
\Biggl [\frac {\Gamma(n+1/2)}{\Gamma(n+1)}\Biggr ]^2z^{-n}
\Biggl [\log(-z)+2\Psi (n+1)-2\Psi(n+1/2)\Biggr ]. \label {sasymp}
\ee
\vspace{0.2cm}

A first approximation $W_0$  to $W$ in (\ref{appix}) is obtained from the
standard relationship between a sum and an integral:

\bea
W_0&=&\frac {L_0}{2\pi}
\int _{\pi/L_0}^{\pi -\pi/L_0} dG(x)
+\frac {1}{4} \biggl [\frac {dG(x)} {dx}\biggl \vert _{x=\frac {\pi}{L_0}} 
                     +\frac {dG(x)} {dx}\biggl \vert _{x=\pi-\frac {\pi}{L_0}}
		     \biggr ] \label {summation} \\
&=&-\frac {L_0}{\pi}G(\frac {\pi} {L_0})+\frac {1}{2}\frac {dG(x)} {dx}\biggl 
\vert
_{x=\frac {\pi}{L_0}},
\eea
where use has been made of the symmetry of $\frac {dG(x)} {dx}$ under 
$x\leftrightarrow \pi-x$. Up to terms which vanish as $\pi/L_0\to 0$,
Eq. (\ref{sasymp}) gives:
\be
W_0\simeq -\frac {1}{2\pi}\biggl [\frac {3\log 2}{2}+1-\log \pi+\log
L_0 \biggr ].
\ee

For that part of $G(x)$ which is analytic at $x=0$, this is the
right answer. A correction to the constant term in $L_0$ however
arises from the logarithmic singularity of $G(x)$   
\bigg (Euler --- Mac-Laurin formula, \cite {abram}\bigg ). As $x\to 0$, we have
$\frac {dG}{dx}\simeq -\frac {1}{\pi}\log \frac {n_0}{L_0}$ up
to analytic terms, whose contribution $W^{log}$ to $W$ from Eq. (\ref
{sasymp}) can be computed using
\be
\sum_{n_0=1}^{L_0/2}\log \frac {n_0}{L_0}=
\log \biggl (\frac {(L_0/2)!}{L_0^{L_0/2} }\biggr )
\simeq \log \frac {\sqrt{\pi L_0}}{(2e) ^{L_0/2}}
\ee
and found to be
\be
W^{log}= -\frac {1}{2\pi}\biggl [\log (2L_0\pi)-L_0(1+\log 2)\biggr ].
\ee
This we compare to the contribution $W_0^{log}$ to $W_0$ of 
the same singular part of $\frac {dG}{dx}$, namely
\be
W_0^{log}=-\frac {1}{2\pi}\biggl [\log (2L_0)-L_0(1+\log
2)\biggr ],
\ee
so that in the limit $L_0\to \infty$ the final result is
\bea
W&=&W_0-W_0^{log}+W^{log} \\
&=&-\frac {1}{2\pi}\biggl [\frac {5}{2}\log 2-1+\log
L_0\biggr ].
\eea

\end{document}